\documentstyle[12pt]{article}
\begin{document}\baselineskip=18pt
\def\be{\begin{equation}}
\def\ee{\end{equation}}
\vskip 1cm
\begin{tabbing}
\hskip 11.5 cm \=\\
\>hep-th/9705243\\
\>May 1997
\end{tabbing}
\vskip 1cm
\begin{center}
{\Large\bf The phases of two-dimensional QED and QCD}
\vskip 1.2cm
{\large \bf  
Roya Mohayaee}\\
\vskip 0.4cm
{{\it International Centre for Theoretical Physics, Trieste, Italy}\\
{{\it Institute of Physics, University of S\~ao Paulo, S\~ao Paulo, Brazil}}\\
\vskip 0.4cm
roya@fma1.if.usp.br\\
mohayaee@ictp.trieste.it}
\end{center}
\abstract 

The semi-classical phase structure of two-dimensional QED and QCD are
briefly reviewed. The non-abelian theory is reformulated to closely 
resemble the Schwinger model. It is shown that, contrary to the
abelian theory, the phase structure of two-dimensional QCD is
unaffected by the structure
of the theta vacuum. We make parallel calculations in the two theories and 
conclude that massless Schwinger model is in the screening and 
the massive theory is in the confining phase, whereas both massless and 
massive QCD are in the screening phase.

\vfill\eject

\section{Introduction}

\indent

Massless Schwinger model is an exactly solvable theory
\cite{elciocuba}. The phase
structure of this theory has been studied extensively. It is
well-established that the theory is in the Higgs or screening
phase. On the other hand, massive Schwinger model is not an exactly
solvable theory. Nevertheless, it has also been studied intensely and
it is known that, under certain approximations, the theory is
confining. There are various ways of establishing the phases
of the Schwinger model. A rather simple method, which we shall demonstrate
here, is to use the bosonised version of the theory and introduce external
probe charges into the system. For a semi-classical theory, the
inter-charge potential binding the test particles can be 
easily computed. Classically, the
Coulomb potential is expected to rise linearly with the
inter-charge separation. However, in the bosonised theory, 
vacuum polarization
effect can shield the probe charges. As a result, in the massless 
Schwinger model, 
the confining Coulomb interaction is replaced by a screening
potential. The massive theory, on the other hand, survives the
polarization effects and is in the confining phase.  

The same method can be applied to two-dimensional 
QCD \cite{elciocuba}. Although 
an exact bosonisation
formulae is not available for the non-abelian theory and the available
bosonisation methods are perturbative in the mass
parameter \cite{rothe}, the techniques developed for Schwinger model
can be used to infer informations about the phase of two-dimensional QCD. 
We introduce external probe
colour charges into the theory and evaluate
their inter-charge potential.
Unlike two-dimensional QED, both the
massless and massive non-abelian theories are in the Higgs phase.

\section{Schwinger model}

\indent

We start with the lagrangian for two-dimensional QED \cite{schwinger},
\be
{\cal L}=-{1\over 4} F_{\mu\nu}F^{\mu\nu} - 
\psi(i\gamma^\mu\partial_\mu-e\gamma^\mu A_\mu - m)\psi.
\ee
This lagrangian can be re-written in terms of the bosonic variables, by
using Mandelstam bosonisation formula \cite{mandelstam}
\begin{eqnarray}
\psi&=&\left(
\begin{array}{c}
\psi_1\\\psi_2
\end{array}
\right)
\\
\psi_1&\sim&:e^{i\beta\phi+{i\over\beta}\tilde\phi}:,\\
\psi_2&\sim&:e^{-i\beta\phi+{i\over\beta}\tilde\phi}: .
\end{eqnarray}
The bosonised lagrangian is
\footnote{Unlike the original lagrangian (1), 
the bosonised lagrangian does not
describe a purely classical system. In
the bosonisation procedure, one takes into account the vacuum
polarization effects by evaluating the contributions
from the one-loop vacuum functionals.} ,
\be
{\cal L}=-{1\over 2}F^{\mu\nu}F_{\mu\nu}+{1\over
2}\partial_\mu\phi\partial^\mu\phi+{\epsilon^{\mu\nu}A_\mu\partial_\nu\phi}
+m^2\gamma({\rm
cos}(2\sqrt\pi\phi)-1),
\ee
where $\gamma$ is a normalisation constant \cite{gross}.
Next, we place external
probe charges $q$ and $-q$ at $L/2$ and $-L/2$. For the
purpose of evaluating the inter-charge potential, it suffices to restrict
ourselves to static fields ({\it i.e.} $\partial_0=0$). 
The static lagrangian, incorporating the probe charges, is 
\begin{eqnarray}
{\cal L}&=&{1\over 2}(\partial_1A_0)^2-{1\over 2}(\partial_1\phi)^2
+m^2\gamma\left[{\rm
cos}(2\sqrt\pi\phi)-1\right]\nonumber\\
&+&{e\over\sqrt\pi}A_0\partial_1\phi+A_0
q\left[\delta(x-L/2)-\delta(x+L/2)\right]
\end{eqnarray}
The equations of the motion  corresponding to the above lagrangian are
\begin{eqnarray}
{e\over\sqrt\pi}\partial_1\phi+q(\delta(x-L/2)-\delta(x+L/2))-
\partial^2_1A_0&=&0\\
-2\sqrt\pi m^2\gamma{\rm sin}(2\sqrt\pi\phi)-{e\over\sqrt\pi}\partial_1A_0
+\partial^2_1\phi&=&0
\end{eqnarray}
The equation of motion of $A_0$ can be integrated to give an
expression for the scalar
field in terms of the electric field $E$ ($\partial_1 A_0$), {\it i.e.},
\be
\phi={\sqrt\pi\over e}[\partial_1A_0-q(T(x-L/2)-T(x+L/2))-\alpha],
\ee
where T is the step function and $\alpha$ is the 
integration constant\footnote{ Note that
in four dimensions, this integration constant is zero. However,
in two dimensions energetics allow for a non-vanishing background
electric field \cite{coleman}.}. This can be inserted into
the equation of motion for $\phi$ to yield,
\be
\partial_1^2\tilde E -{e^2\over\pi}\tilde E - {eq\over\sqrt\pi}
(T-T)-2\sqrt\pi m^2\gamma{\rm sin}(2\sqrt\pi \tilde
E-{2\pi\alpha\over e})=0
\ee
where 
\be
\tilde E={\sqrt\pi\over e}[E-q(T-T)],
\ee
$\theta=2\pi\alpha/e$ is the theta vacuum and
$(T-T)$ denotes $(T(x-L/2)-T(x+L/2))$.

In order to obtain the potential binding $q$ and $-q$, 
we solve the above
equation to first find the inter-charge electric field. The 
equation can be solved for two different cases: when the
dynamical fermions are massless and when they are massive.

\subsection{Massless Schwinger model}

\indent

For massless dynamical fermions, the equation of motion for $\tilde E$
(10) reduces to
\be
\partial^2\tilde E-{e^2\over\pi}\tilde E -{e\over\sqrt\pi} q (T-T)=0.
\ee
This equation is exactly solvable and its solutions are
\begin{eqnarray}
\tilde E_I&=& a \exp({{-e\over \sqrt\pi}x}),\qquad\quad
\qquad\quad\qquad\quad\quad x> {L\over 2}\\
\tilde E_{II}&=& b\exp({{e\over \sqrt\pi}x}),\qquad\quad
\qquad\quad\qquad\quad\quad x< {-L\over 2}\\
\tilde E_{III}&=& c\exp({{-e\over \sqrt\pi}x})+d\exp({{e\over
\sqrt\pi}x})+{\sqrt\pi q\over e},
\quad{-L\over 2} <x< {L\over 2}
\end{eqnarray}
The solutions and their derivatives can be matched at the boundaries
to easily obtain the unknown coefficients. Having obtained $\tilde E$, we use
the expression (11) to obtain the electric field. The
inter-charge potential $V(L)$ can then be simply obtained by integrating
the electric field over the inter-charge separation \footnote{This can
be shown to be equivalent to evaluating the change in the hamiltonian
caused by the probe charges (see \cite{elcio}, Chapter 10).}, 
{\it i.e.},
\be
V(L)=-q\int^{L/2}_{-L/2} E_{III} dx={q^2\sqrt\pi\over 2e}(1-e^{-eL/\sqrt\pi})
\ee

The expression for the inter-charge potential shows that
for small inter-charge separations,
the potential rises linearly with $L$, as is expected from a classical
coloumb potential. However, as the inter-charge separation is
increased the polarization effects set up \footnote{Recall that we have
taken into account the polarization effects in the bosonisation
scheme. In replacing the original fermionic lagrangian by the bosonic
lagrangian, one incorporates the contributions from the one-loop 
Feynman diagrams in the bosonic action.} and lead to the screening of
the probe charges. For very large inter-charge separations, the potential
eventually reaches the constant value $q^2\sqrt\pi /2e$. 

So far, the charge of the probe particles has not been specified,
Therefore, even the screening of the 
fractional probe charges by
integer dynamical charges is allowed. This is puzzling since 
one would expect the
confinement to prevail in such cases. 
The situation can be clarified by evaluating
the conserved charge $Q$ associated with the integer
dynamical charges. This charge is given in terms of the conserved
current $\partial_1\phi$, {\it i.e.},
\be
Q=\int^{x_2}_{x_1} dx \partial_1\phi.
\ee
To unravel the mechanism of screening, we consider the screening of one of
the probe charges and evaluate the conserved charge
along the axis from 0 to $\infty$ ({\it i.e.}, a solitonic
configuration). That is,
\be
Q=\phi(\infty)-\phi(0)=-q.
\ee

This shows that although the dynamical charges are of 
integer values, the charge
associated with their solitonic
configuration can be non-integer. Specifically, this charge is
opposite to the charge of the probe particles and accounts for the 
shielding phenomena.

\subsection{Massive Schwinger model}

\indent

For massive dynamical fermions, the Schwinger model is not
exactly solvable. The equation of motion (10) is
non-linear and can only be solved
after expanding the sine-term \footnote{The validity of this
approximation can be checked by evaluating the argument of the
sine, using the approximate solution for the electric field. One
verifies that, for a
large mass to charge ratio, the argument of the sine is much smaller
than $\pi/4$. Thus, the expansion of the sine term is only meaningful for
$m>> e$.}. In this approximation, the equation 
of motion (10) reduces to
\be
\partial_1\tilde E-({e^2\over\pi}+4\pi m^2\gamma)\tilde E -
({e\over\sqrt\pi}q(T-T) + 2\sqrt\pi m^2\gamma\theta)=0.
\ee

This equation resembles (12) for the massless Schwinger
model and can be solved in the same manner. By using the
expression (11) relating $\tilde E$ to the electric field, we
obtain the electric field and subsequently the inter-charge potential.
This is now given by
\begin{eqnarray}
V(L)&=&{e^4\over 2\pi(e^2/\pi+4\pi m^2\gamma)^{3/2}}
(1-e^{-\sqrt{(e^2/\pi+4\pi m^2\gamma)}L})\nonumber\\
&+&
{e\over2}(1-{e^2\over e^2+4\pi^2 m^2\gamma})(q-{e\theta\over\pi})L
\end{eqnarray}
Therefore, for the massive fermions, the inter-charge potential has
both a screening and a confining term. However, for long separations
the confinement term dominates. In addition, for integer
probe charges and for $\theta=\pi$ a phase transition occurs; the
confinement term disappears and the screening phase is restored. This
can be explained by recalling that the theta vacuum, which
was introduced in equation (9) as an integration constant, is basically
a non-vanishing background electric field. 
As the theta angle is increased, pair production 
sets up helping the screening of the probe charges. This continues until
the net electric field falls below the threshold required for pair
production. This circle repeats itself and therefore the dynamics of the
system is a periodic function of the theta angle.

\subsection{Two-dimensional QCD}

\indent

In the preceding sections, we have studied the screening and confining
phases in two-dimensional QED. We saw that although massless Schwinger
model is in the screening phase, the massive theory exhibit
confinement. In this section, we ask the same question for
two dimensional QCD and examine whether the conclusions drawn for
two-dimensional QED
can be generalized to its non-abelian counterpart. 

The action of two-dimensional QCD is

\be
S=\int d^2 x \left[-{1\over 4}{\rm tr}F_{\mu\nu}F^{\mu\nu}+
\bar\psi_i^f(i{\not\!\partial}
\delta^{ij}-e{\not\!\! A}^{ij})\psi^f_j-m
\delta^{ij}\bar\psi^f_i\psi^f_j\right],
\ee
where $i,j$ are the usual colour indices and $f=1,\cdots ,k$ is a flavour 
quantum number. Unlike the Schwinger model, for which an exact
bosonisation scheme (3) exists, two-dimensional QCD cannot be exactly
bosonised. However a perturbative bosonisation scheme by means of which the
massless theory is first bosonised and the mass term is
introduced and bosonised perturbatively exists \cite{rothe}. 
Writing down the generating function for the above lagrangian 
and integrating out the fermions in the
path integral measure leads to the well-known Wess-Zumino-Witten action
\begin{eqnarray}
S_{{\rm eff}}&=&\sum_f\Gamma[g_f]-(c_v+k)\Gamma[\Sigma]
+k\Gamma[\beta]+S_{\rm YM}\nonumber\\
&+&m^2\int d^2 z{\rm tr}\left[\sum_f
(g_f\Sigma^{-1}\beta+g^{-1}_f\Sigma\beta^{-1})\right],
\end{eqnarray}
where
\begin{eqnarray}
S_{\rm YM}&=&\int d^2z\left[{1\over 2}(\partial_+ A_0)^2+\lambda A_0
(\beta^{-1}i\partial_+\beta)\right],\\
\Gamma[g]&=&\int{d^2x\over 8\pi} {\rm tr}\left(\partial_\mu
g^{-1}\partial^\mu g\right) +\int{d^3y\over 12\pi}
\epsilon^{\alpha\beta\gamma} {\rm tr}\left(g^{-1}\partial_\alpha g
g^{-1}\partial_\beta g g^{-1} \partial_\gamma g\right),\nonumber\\
& &
\end{eqnarray}
the light-cone coordinate is $x^+$, 
$\lambda={e\over 2\pi}(c_v+k)$ and  $c_v$ is given 
by $c_v f^{ad}=f_{abc}f^{cbd}$ which vanishes for the abelian
group \footnote{We shall make frequent use of this limit and make
parallels with the Schwinger model by taking
the limit $c_v\rightarrow 0$.}. The $g$ field is the
gauge-invariant bosonic field corresponding to the original fermionic
excitations. The field $\Sigma$ is a negative-metric field and the
fields $\beta$ and $A_0$ are the massive-sector fields. 
The equations of motion corresponding to this action are
\begin{eqnarray}
{1\over 4\pi}\partial_+(g_f\partial_-g_f^{-1})&=&m^2(g_f\Sigma^{-1}\beta -
\beta^{-1}\Sigma g^{-1}_f),\quad f=1\cdots k,\nonumber\\
& &\\
-{(c_v+k)\over 4\pi}\partial_+(\Sigma\partial_-\Sigma^{-1})&=&
m^2\sum_{f=1}^k(\Sigma g_f^{-1}\beta^{-1}-\beta g_f\Sigma^{-1}) ,\\
-{k\over 
4\pi}\partial_-(\beta^{-1}\partial_+\beta)&+&i\lambda[\beta^{-1}\partial_+
\beta ,A_0]+i\lambda\partial_+ 
A_0 \, \nonumber\\
&=&m^2\sum_{f=1}^k(g_f\Sigma^{-1}
\beta-\beta^{-1}\Sigma g_f),\\
\partial_+^2 A_0&=&\lambda(\beta^{-1}i\partial_+\beta).
\end{eqnarray}

Unlike the Schwinger model, where the equations of motion of 
the bosonised theory, (7) and (8), were
given in terms of simple scalar fields, all the above fields are
matrices. To obtain a set of solvable equations, we parametrise the
matrix-valued fields and rewrite them as elements of the gauge
group. That is,
\be
g= e^{i\sigma_2\varphi}\,\,\,\,\, ,\,\,\Sigma=e^{-i\sigma_2\eta}\,\,\,\,\,
,\,\,\beta=e^{i\sigma_2\zeta} \,\ ,
\ee
where the fields $\varphi,\eta$ and $\zeta$ are scalars and
 $\sigma_2$ is a generator of SU(2) group \footnote{ This
simple parametrisation can be easily
extended to SU(N). }. Rewriting the lagrangian (22) in terms of the
above variables, introducing external colour probe charges
$q^a$, with fixed colour charges $a$,
and taking the static limit
($\partial_0=0$), we obtain the equations of motion
\begin{eqnarray}
\partial^2\varphi&=& 8\pi m^2{\rm sin}(\varphi+\eta+\zeta),\\
\partial^2\eta&=&{-8\pi m^2\over (c_v+1)}{\rm
sin}(\varphi+\eta+\zeta),\\
\partial E&=&\partial^2 A_0={(c_v+1)\over 2\pi}q^a(\delta(x-L/2)-
\delta(x+L/2))-\lambda\partial\zeta,\\
E&=&{-1\over 4\pi\lambda}\partial^2\zeta +{2m^2\over\lambda}{\rm
sin}(\varphi+\eta+\zeta)-
\end{eqnarray}
The integration constant $\alpha$ arising from the integration of 
equation (32) can be interpreted as the background electric field 
{\it i.e.}, as the theta vacuum; $\theta=(2\pi\alpha/e)$ (see equations
(9) and (10) of the Schwinger model for comparison.).

By making the substitution $\phi=\varphi+\eta$, the above four
equations are reduced to the coupled equations,
\begin{eqnarray}
\partial^2\tilde E&=&4\pi\lambda^2\tilde E+8\pi m^2{\rm
sin}(\phi+\tilde E+{\alpha\over\lambda})\nonumber\\
&-&2\lambda q^a
(c_v+1)(T(x-L/2)-T(x+L/2)),\\
\partial^2\phi&=&({8\pi m^2 c_v\over c_v+1}){\rm sin}(\phi+\tilde
E+{\alpha\over\lambda}),
\end{eqnarray}
where $\tilde E=\zeta-\alpha/\lambda$ and
$E=-\lambda\tilde E+{c_v+1\over 2\pi}q^a(T-T)$ (similar to (11) 
of the Schwinger model).

\section{Massless QCD}

\indent

For massless dynamical fermions, equation (34)
simplifies to
\be
\partial^2\tilde E-4\pi\lambda^2\tilde E
+2q^a\lambda(c_v+1)(T(x-L/2)-T(x+L/2))=0.
\ee
This equation is similar to the expression (12) of the Schwinger
model and can be solved by the same techniques. The 
inter-charge potential, obtained in the fashion of the Schwinger
model, is
\be
V(L)={(c_v+1)\sqrt\pi {q^a}^2\over 2 e}(1-e^{-(c_v+1)eL\over\sqrt\pi}).
\ee
Thus, massless QCD exhibits screening. It is worth mentioning that 
the screening potential of the Schwinger model (16)
can be easily obtained by taking the limit $c_v\rightarrow 0$.

\section {Massive two-dimensional QCD}

\indent

The massive equations of motion, (34) and (35),
are not exactly solvable. We first
expand the sine term and solve 
the coupled equations for the field $\phi$. The quartic 
equation for the field $\phi$ is
\begin{eqnarray}
\partial^4\phi&-&\left({8\pi m^2c_v\over c_v+1}+4\pi\lambda^2+8\pi
m^2\right)\partial^2\phi
-{32\pi^2\lambda^2m^2 c_v\over c_v+1}\phi\nonumber\\
&-&\left({32\pi^2\lambda c_v\over c_v+1}\alpha+
16\pi m^2 c_v\lambda q^a(T-T)\right)=0.
\end{eqnarray}
The electric field can be obtained from the solutions of the above
equation by using expression (35) and the relation between $E$ and
$\tilde E$. Subsequently, the inter-charge
potential is
\begin{eqnarray}
& &V(L)={(c_v+1)^2{q^a}^2\over 2}\nonumber\\
& &\quad\times\left[ 
\!\!\left({4\pi\lambda^2-m_-^2\over 
m_+^2-m_-^2}\right)\!\!\!\left({1-e^{-m_+L}\over m_+}\right)
\!+\!\left({m_+^2-4\pi\lambda^2\over 
m_+^2-m_-^2}\right)\!\!\!\left({1-e^{-m_-L}\over 
m_-}\right)\!\!\right],\nonumber\\ & & 
\end{eqnarray}
where the mass scales $m_{\pm}$, arising from the solutions of the quartic
equation (38), are given by
\vfill\eject
\begin{eqnarray}
m^2_\pm&=&2\pi\left[\lambda^2+
\left(1+{c_v\over (c_v+1)}\right)2  m^2\right]
\nonumber\\
&\pm&2\pi\left[\sqrt{\left(\lambda^2+(1+{c_v\over (c_v+1)})2m^2\right)^2-
8{c_v\over (c_v+1)}\lambda^2 m^2}\right].
\end{eqnarray}

The expression (39) for the inter-charge potential contains no
confining term. Thus, massive QCD is in the screening phase. The
confining potential of the massive Schwinger model is obtained 
by taking the limit
$c_v\rightarrow 0$. In this limit, the mass scale $m_-$ goes to zero
and we recover expression (20) of the Schwinger model for $\theta=0$.
We also observe that, unlike the Schwinger model, the theta vacuum,
{\it i.e.,} 
the constant $\alpha$ in (35), 
does not appear in the expression
for the inter-charge potential. Thus, the phase structure of the
non-abelian theory is not affected by the values of the theta angle.  

\section{Conclusion}

\indent

By a simple semi-classical treatment of two dimensional QED and QCD,
we have determined the phases of these theories. The
introduction of the mass parameter in the Schwinger model
causes a transition
from the screening to the confining phase. This transition does
not occur in the non-abelian theory where
the screening phase prevails. It is worth mentioning that similar
analysis were recently done in higher dimensions 
\cite{elcioandbanerjee}. It has been shown 
that both massless and massive (for very large fermion masses) 
three-dimensional QED are in the screening phase \cite{elcioandbanerjee}.

\section{Acknowledgements}
\indent

This article is a simpler presentation of the results
obtained in collaboration with E.
Abdalla and A. Zadra\cite{ayrton}. 
I would like to thank E. Abdalla for a critical
reading of the manuscript and numerous helpful discussions. This work
was done under financial support from Funda\c c\~ao de Amparo a
Pesquisa do Estado de S\~ao Paulo (FAPESP).


\end{document}